\documentclass[aps,pre,floatfix]{revtex4}
\usepackage{epsf}
\usepackage{subfigure}
\usepackage{amsmath}
\usepackage{amssymb}
\usepackage{graphicx}
\usepackage{epstopdf}
\font\helvb=cmssbx12

\begin{document}

\newcommand{\be}{\begin{equation}}
\newcommand{\ee}{\end{equation}}
\newcommand{\bea}{\begin{eqnarray}}
\newcommand{\eea}{\end{eqnarray}}
\newcommand{\mean}[1]{\left \langle #1 \right \rangle}

\title{\bf Time-reversal symmetry relation for nonequilibrium flows \\ ruled by the fluctuating Boltzmann equation}

\author{Pierre Gaspard}
\affiliation{Center for Nonlinear Phenomena and Complex Systems \& Department of Physics,\\
Universit\'e Libre de Bruxelles, Code Postal 231, Campus Plaine, 
B-1050 Brussels, Belgium}

\begin{abstract}
A time-reversal symmetry relation is established for out-of-equilibrium dilute or rarefied gases described by the fluctuating Boltzmann equation. The relation is obtained from the associated coarse-grained master equation ruling the random numbers of particles in cells of given position and velocity in the single-particle phase space. The symmetry relation concerns the fluctuating particle and energy currents of the gas flowing between reservoirs or thermalizing surfaces at given particle densities or temperatures.  


\end{abstract}

\noindent {}

\vskip 0.5 cm

\maketitle

\section{Introduction}
\label{Intro}

Since 1872, Boltzmann's equation has provided the main paradigm of our understanding of irreversible phenomena~\cite{B72}. In isolated systems such as a dilute gas in a container, the $H$-theorem established by Boltzmann with his equation shows that any velocity distribution for the particles irreversibly converges at long time towards the Maxwell velocity distribution characterizing the thermodynamic equilibrium \cite{B95}. This relaxation towards equilibrium is generated by successive binary collisions between the particles composing the gas.  They are described as in chemical kinetics by the mass action law, requiring that the rate of binary collisions be proportional to the concentrations of particles of the corresponding velocities in the volume element where the collisions happen.  For this reason, Boltzmann's equation is nonlinear as the mean-field kinetic equations describing macroscopic chemical reactions, except that the particle velocities are not macroscopic observables \cite{N72}.  Nevertheless, finer observables closer to the microscopic level of description and, especially, fluctuations remain outside the framework of Boltzmann's theory.

Since the forties, a fluctuating Boltzmann equation has been proposed which rules the local velocity distribution function as a random variable, much in analogy with Langevin's stochastic equation \cite{S49,BZ69,FU70,vK74,LK76,O78,BVdB80,EC81,vK81,G04}.  This formulation provides a description closer to the microscopic level and, thus, more suitable to understand the properties of the fluctuations. Although they are known to be time-reversal symmetric at equilibrium because of the principle of detailed balancing, few results are available about fluctuations in nonequilibrium steady states.

\begin{figure}[h]
\begin{center}
\includegraphics[scale=0.43]{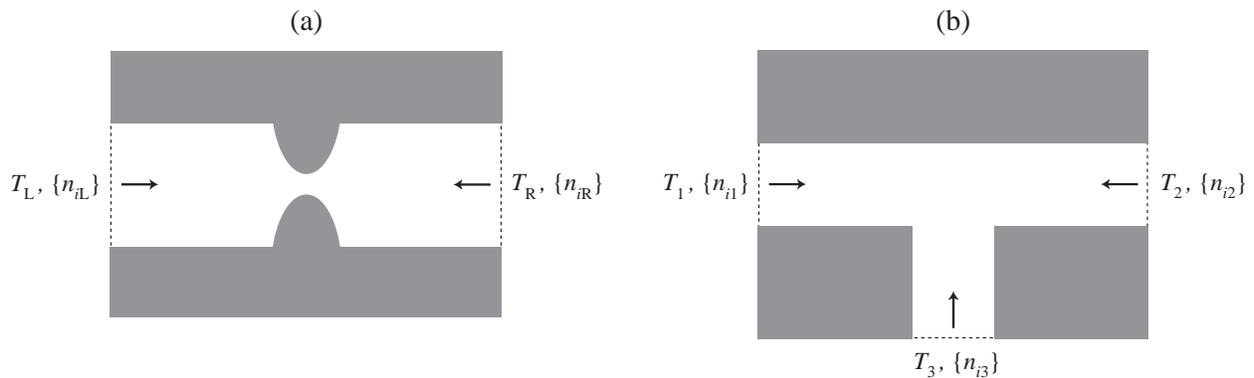}
\caption{Schematic representation of dilute or rarefied gases flowing through pores or pipes between reservoirs at given temperatures and particle densities.  In gas mixtures, there are several particle species $i$ and so many particle densities.}
\label{fig1}
\end{center}
\end{figure}

The purpose of the present paper is to establish a time-reversal symmetry relation valid out of equilibrium for the fluctuating Boltzmann equation.  Such relations have been the subject of the so-called fluctuation theorems, which have been obtained in particular for stochastic processes such as diffusion processes ruled by Langevin's equations and their associate Fokker-Planck equation or continuous-time Markovian jump processes \cite{ECM93,GC95,K98,LS99}. In general, a fluctuation theorem holds for all the currents flowing across an open system in nonequilibrium steady states \cite{AG07JSP}.  Such a theorem implies the second law of thermodynamics and it allows us to deduce generalizations of the Green-Kubo formulae and Onsager reciprocity relations for the nonlinear response properties \cite{AG07JSM,GA11}.  Our aim is here to extend these fundamental results to the dilute and rarefied gases ruled by Boltzmann's equation \cite{K1909,P58,C00}.  We consider such gases flowing under nonequilibrium conditions in pores, pipes, or other ducts between several reservoirs, as illustrated in Fig.~\ref{fig1}.

The paper is organized as follows. In section \ref{Boltzmann}, Boltzmann's equation is introduced as a mean-field equation in open geometries with gas-surface interactions. The symmetries of the collision kernel and, in particular, its time-reversal symmetry are discussed.  In section \ref{master_eq}, the one-particle phase space is partitioned into cells and the coarse-grained master equation is obtained for the probability that the cells contained certain particle numbers. In section \ref{fluctuating_Boltzmann}, the diffusive approximation of the coarse-grained master equation is shown to lead to the fluctuating Boltzmann equation including the terms due to the contacts with the reservoirs. The time-reversal symmetry relation is proved in section \ref{TRsym} where the corresponding fluctuation theorem is established.  Conclusions are drawn in section \ref{conclusions}.

\section{Boltzmann's equation as a mean-field kinetic equation}
\label{Boltzmann}

At the microscopic level of description, the motion of the particles composing the gas is ruled by Newton's equations for their positions and velocities $\{{\bf r}_n(t),{\bf v}_n(t)\}$.  The particle distribution function is defined as the density to find one particle with the position $\bf r$ and the velocity $\bf v$ at the current time $t$:
\be
f({\bf r},{\bf v},t) \equiv \sum_n \delta\left[ {\bf r}-{\bf r}_n(t)\right] \, \delta\left[ {\bf v}-{\bf v}_n(t)\right]
\label{f}
\ee
for monoatomic particles without internal rotation or vibration.
Given that the initial conditions of the particles are distributed according to some probability distribution for the whole system, the particle distribution function (\ref{f}) is a random variable which may fluctuate. We may also consider its average value over the given probability distribution:
\be
\langle f({\bf r},{\bf v},t)\rangle \equiv \sum_n \langle \delta\left[ {\bf r}-{\bf r}_n(t)\right] \, \delta\left[ {\bf v}-{\bf v}_n(t)\right]\rangle
\label{<f>}
\ee
which is expected to smoothly vary in space and time.

\subsection{In the bulk of the flow}

For a general classical system, the time evolution of the one-particle average distribution function (\ref{<f>}) is ruled by the first equation of the Bogoliubov-Born-Green-Kirkwood-Yvon hierarchy \cite{B75,RL77,H87,P09}.  The other equations of this hierarchy are ruling the many-particle average distribution functions.  In the dilute-gas limit, the many-particle distribution functions factorize in terms of the one-particle function.  The time evolution of the one-particle function is determined by the binary collisions and their differential cross section, which is obtained using classical scattering theory.  Supposing that the particles are incoming every binary collision without statistical correlation as stated in the famous ``Stosszahlansatz", Boltzmann's equation is deduced, which is a closed equation for the one-particle average distribution function (\ref{<f>}) of the following form:
\be
\frac{\partial\langle f\rangle}{\partial t} + {\bf v}\cdot\frac{\partial\langle f\rangle}{\partial{\bf r}} 
+ \frac{\bf F}{m}\cdot\frac{\partial\langle f\rangle}{\partial{\bf v}} = \int d{\bf v}_2 \; d{\bf v}_1' \; d{\bf v}_2' 
\; w\left({\bf v}_1,{\bf v}_2\vert{\bf v}_1',{\bf v}_2'\right) \; \left(\langle f_1'\rangle\, \langle f_2'\rangle-\langle f_1\rangle\, \langle f_2\rangle\right) 
\label{B_eq}
\ee
where $\langle f\rangle=\langle f_1\rangle$ is the average distribution function at the position ${\bf r}={\bf r}_1$ and the velocity ${\bf v}={\bf v}_1$ of the first particle involved in the binary collision \cite{B75,RL77,H87,P09}.  ${\bf F}={\bf F}({\bf r})$ is an external force field, which includes the repulsive forces of the walls of the duct.  In general, the transition rates have the following symmetries:
\bea
\mbox{time-reversal symmetry:}\qquad && w\left({\bf v}_1,{\bf v}_2\vert{\bf v}_1',{\bf v}_2'\right) = w\left(-{\bf v}_1',-{\bf v}_2'\vert -{\bf v}_1,-{\bf v}_2\right) \label{T-sym}\\
\mbox{space-orthogonal symmetry:}\qquad && w\left({\bf v}_1,{\bf v}_2\vert{\bf v}_1',{\bf v}_2'\right) = w\left(\mbox{\helvb R}\cdot{\bf v}_1,\mbox{\helvb R}\cdot{\bf v}_2\big\vert \mbox{\helvb R}\cdot{\bf v}_1',\mbox{\helvb R}\cdot{\bf v}_2'\right) \label{R-sym}
\eea
where $\mbox{\helvb R}$ is a matrix belonging to the orthogonal group O(3) including spatial rotations and reflections \cite{H87,P09}.  We also have the
\be
\mbox{inversion symmetry:}\qquad w\left({\bf v}_1,{\bf v}_2\vert{\bf v}_1',{\bf v}_2'\right) = w\left({\bf v}_1',{\bf v}_2'\vert {\bf v}_1,{\bf v}_2\right) \label{PT-sym}
\ee
which is the symmetry under the time-reversal transformation combined with the spatial inversion of orthogonal matrix $\mbox{\helvb R}=-\mbox{\helvb I}$ where $\mbox{\helvb I}$ is the identity matrix.

The transition rates are related to the differential cross section $\sigma_{\rm diff}$ of the binary collisions as
\be
w\left({\bf v}_1,{\bf v}_2\vert{\bf v}_1',{\bf v}_2'\right) = \sigma_{\rm diff} \; \delta\left({\bf v}_1+{\bf v}_2-{\bf v}_1'-{\bf v}_2'\right) \; \delta\left({\bf v}_1^2+{\bf v}_2^2-{\bf v}_1'^2-{\bf v}_2'^2\right)
\ee
where the delta's express the conservation of linear momentum and kinetic energy in every binary collision.  Setting ${\bf V}'=({\bf v}_1'+{\bf v}_2')/2$ and ${\bf u}'={\bf v}_1'-{\bf v}_2'$ and integrating over the velocities $d{\bf v}_1' d{\bf v}_2' = d{\bf V}'d{\bf u}'=d{\bf V}' u'^2 du'd\Omega$, Boltzmann's equation (\ref{B_eq}) takes its more usual form:
\be
\frac{\partial\langle f\rangle}{\partial t} + {\bf v}\cdot\frac{\partial\langle f\rangle}{\partial{\bf r}} 
+ \frac{\bf F}{m}\cdot\frac{\partial\langle f\rangle}{\partial{\bf v}} = \int d{\bf v}_2 \; d\Omega \;
\sigma_{\rm diff} \; \Vert{\bf v}_1-{\bf v}_2\Vert \; \left(\langle f_1'\rangle\, \langle f_2'\rangle-\langle f_1\rangle\, \langle f_2\rangle\right)
\label{B_eq_usual}
\ee
in terms of the differential cross section and the solid angle element $d\Omega$.  For a gas of hard spheres of radius $a$, this cross section is given by $\sigma_{\rm diff}=a^2$.

In the absence of force field ${\bf F}=0$, Boltzmann's equation is known to admit stationary solutions given by the overall equilibrium Maxwell-Boltzmann distributions:
\be
f_{\rm eq}({\bf v}) = \left(\frac{m\beta}{2\pi}\right)^{3/2} n \; \exp(-\beta\epsilon)
\qquad \mbox{with}\qquad \epsilon = \frac{1}{2}m{\bf v}^2
\label{MB_eq}
\ee
for particles of mass $m$ at the uniform temperature $T=(k_{\rm B}\beta)^{-1}$ and density $n$.  
Since the transition rates preserve the conservations of linear momentum and kinetic energy in the binary collisions, the time-reversal symmetry (\ref{T-sym}) also implies the property of reciprocity:
\be
w\left({\bf v}_1,{\bf v}_2\vert{\bf v}_1',{\bf v}_2'\right)\; f_{\rm eq}({\bf v}'_1) \; f_{\rm eq}({\bf v}'_2) = w\left(-{\bf v}_1',-{\bf v}_2'\vert -{\bf v}_1,-{\bf v}_2\right)\; f_{\rm eq}({\bf v}_1) \; f_{\rm eq}({\bf v}_2)  \label{reciprocity-sym}
\ee
We notice that the same property still holds in an inertial frame moving at the velocity ${\bf V}_0$ where the equilibrium distribution is given by Eq.~(\ref{MB_eq}) with $\bf v$ replaced by ${\bf v}-{\bf V}_0$.  Moreover,
the property of reciprocity as well as the stationarity of Boltzmann's equation for the Maxwell-Boltzmann distribution hold locally at every space point in the system.  These two features constitute the basis of what is called the local thermodynamic equilibrium in this system~\cite{B75}.

In the following, the form (\ref{B_eq}) of Boltzmann's equation will be used because its symmetries (\ref{T-sym}) and (\ref{R-sym}) are more explicit.

\subsection{At the boundaries}

For open systems such as those depicted in Fig.~\ref{fig1}, there are several types of boundaries.  

\subsubsection{At the reservoirs}

First, there are the inlets from the reservoirs, by which gas is also evacuated in the opposite directions.  The contacts with the reservoirs contribute to Eq.~(\ref{B_eq}) by terms of the form:
\be
\frac{\partial\langle f\rangle}{\partial t}\Big\vert^{\rm R} = \sum_s \int_s d^2S \; \delta({\bf r}-{\bf r}_s) \; ({\bf n}\cdot{\bf v}) \; \theta({\bf n}\cdot{\bf v}) \; \left[ f_s({\bf v})-\langle f({\bf r},{\bf v};t)\rangle\right]
\label{B_eq_R}
\ee
where $\theta(x)$ is Heaviside's function, ${\bf r}_s$ is the position of a point on the surface $s$ separating the system from the corresponding reservoir, $\bf n$ is a unit vector normal to the surface $s$ and directed inside the system and 
\be
f_s({\bf v}) = \left(\frac{m\beta_s}{2\pi}\right)^{3/2} n_s \; \exp(-\beta_s\,\epsilon)
\qquad \mbox{with}\qquad \epsilon = \frac{1}{2}m{\bf v}^2
\label{MB_distrib}
\ee
is the Maxwell-Boltzmann distribution function of the gas incoming through the surface $s$ from the reservoir at the temperature $T_s=(k_{\rm B}\beta_s)^{-1}$ and the particle density $n_s$. The contacts with the reservoirs are carried out at surfaces which are assumed to be far enough in the pipes for the flow profile to be independent of the details of the boundary conditions.  

These inlets drive the system out of equilibrium if the reservoirs are at different temperatures or particle densities.  For instance, in the case of the system of Fig.~\ref{fig1}a with two reservoirs $s={\rm L},{\rm R}$ and one species of particles, there are two possible differences and thus two thermodynamic forces also called affinities which are the control parameters of the nonequilibrium constraints.  These affinities are defined as
\bea
&&\mbox{thermal affinity:} \qquad \ A_E = \beta_{\rm R}-\beta_{\rm L} \label{A_E}\\
&&\mbox{chemical affinity:} \qquad A_N = \beta_{\rm L}\mu_{\rm L}-\beta_{\rm R}\mu_{\rm R}=\ln\frac{n_{\rm L}\beta_{\rm L}^{3/2}}{n_{\rm R}\beta_{\rm R}^{3/2}} \label{A_N}
\eea
where $\mu_{\rm L,R}$ are the chemical potentials of the monoatomic gas in the left- and right-hand reservoirs. We notice that more than two affinities are defined for an open system in contact with more than two reservoirs at different temperatures or particle densities as illustrated in Fig.~\ref{fig1}b.
If all the affinities are vanishing, the system is at equilibrium and Boltzmann's equation admits the corresponding Maxwell-Boltzmann distribution as stationary solution.  

\subsubsection{At the surfaces}

Now, we consider the interactions of the gas with solid surfaces, on which the particles are scattered. In the simplest model, the gas particles undergo elastic collisions on the surface, so that the outgoing velocity is given by
\be
{\bf v}'={\bf v} - 2 ({\bf n}\cdot{\bf v}) \, {\bf n}
\label{specular}
\ee
where $\bf n$ is a unit vector normal to the surface at the point of collision.  Accordingly, the velocity component perpendicular to the surface changes its sign: ${\bf n}\cdot{\bf v}'=-{\bf n}\cdot{\bf v}$. Moreover, the kinetic energy is conserved because ${\bf v}'^2={\bf v}^2$.  Therefore, the contact of the gas with the surface does not change its temperature or density, as if the gas was in a force field ${\bf F}({\bf r})$ that is very steep at the surface.  In this case, the contribution of the surface to Eq.~(\ref{B_eq}) would be of the form
\be
\frac{\partial\langle f\rangle}{\partial t}\Big\vert^{\rm S} = \int_s d^2S \; \delta({\bf r}-{\bf r}_s) \; ({\bf n}\cdot{\bf v}) \; \theta({\bf n}\cdot{\bf v}) \; \left\{ \langle f[{\bf r},{\bf v}- 2({\bf n}\cdot{\bf v})\,{\bf n};t]\rangle - \langle f({\bf r},{\bf v};t)\rangle \right\}
\label{S-elastic}
\ee

In general, the interaction of the gas with the surface of a solid depends on many different aspects.  The scattering process of particles with surfaces have been much studied and many processes are known besides elastic collisions: adsorption, desorption, transport on the surface, transport into or from the bulk of the solid, or possible reactions~\cite{KG86}.  The solid surface is typically at some temperature that may differ from the temperature of the gas and a transfer of energy can happen at the surface or inside the solid.  Here, the solid forming the container is supposed to have a high enough thermal conductivity so that its temperature is uniform.  Under certain thermodynamic conditions of pressure and temperature, gas particles may diffuse into the solid and form a stable thermodynamic phase such as solid hydrides, oxides, or nanoporous composites.  Under nonequilibrium conditions, the solid could thus become a sink or a source of gas particles, in which case the difference of chemical potentials of the particles between the gas and the solid would add an extra affinity.  For simplicity, this situation is not considered here and only thermal exchanges between the gas and the solid are envisaged.  Surface adsorption and desorption may also occur but the surface is supposed to be in local thermodynamic equilibrium with the gas so that the thermodynamic state of the surface (in particular, the particle coverages) is thus stationary.  Under such circumstances, the contribution of the surface to Eq.~(\ref{B_eq}) would be of the form
\be
\frac{\partial\langle f\rangle}{\partial t}\Big\vert^{\rm S} = \int_s d^2S \; \delta({\bf r}-{\bf r}_s) \left[ \int_{{\bf n}\cdot{\bf v}'<0} d{\bf v}' \vert{\bf n}\cdot{\bf v}'\vert \; p_{\bf r}({\bf v}\vert{\bf v}') \; \langle f({\bf r},{\bf v}';t)\rangle - ({\bf n}\cdot{\bf v}) \; \theta({\bf n}\cdot{\bf v}) \; \langle f({\bf r},{\bf v};t)\rangle\right]
\label{term-S}
\ee
where $p_{\bf r}({\bf v}\vert{\bf v}')$ is the probability density that a particle impinging the surface at the position $\bf r$ with the velocity ${\bf v}'$ and ${\bf n}\cdot{\bf v}'<0$ will be scattered to the velocity $\bf v$ such that ${\bf n}\cdot{\bf v}>0$ \cite{C00}.  We notice that the probability density does not depend on the position $\bf r$ if the surface is homogeneous and the scattering process is the same everywhere.
In general, this function is normalized according to
\be
\int_{{\bf n}\cdot{\bf v} >0} p_{\bf r}({\bf v}\vert{\bf v}') \; d{\bf v} = 1 \qquad\mbox{if}\qquad {\bf n}\cdot{\bf v}'<0
\label{p-norm}
\ee
and it satisfies the following two properties.  The first one guarantees the preservation of equilibrium at the temperature $T_s$ of the surface:
\be
\vert{\bf n}\cdot{\bf v}\vert \; f_s({\bf v}) = \int_{{\bf n}\cdot{\bf v}'<0} p_{\bf r}({\bf v}\vert{\bf v}') \; \vert{\bf n}\cdot{\bf v}'\vert \; f_s({\bf v}') \; d{\bf v}' 
\label{p-eq}
\ee
where $f_s({\bf v})$ is the Maxwell-Boltzmann equilibrium distribution (\ref{MB_distrib}) at the temperature of the wall.  The second is the property of reciprocity:
\be
\vert{\bf n}\cdot{\bf v}'\vert \; f_s({\bf v}') \; p_{\bf r}({\bf v}\vert{\bf v}') = \vert{\bf n}\cdot{\bf v}\vert \; f_s({\bf v}) \; p_{\bf r}(-{\bf v}'\vert -{\bf v})
\label{p-reciprocity}
\ee
which is implied by the time-reversal symmetry of the underlying microscopic dynamics and the condition of local thermodynamic equilibrium of the surface at the temperature $T_s$ of the Maxwell-Boltzmann equilibrium distribution~$f_s({\bf v})$~\cite{C00}.

The special case (\ref{S-elastic}) of elastic collision is recovered for
\be
p_{\bf r}({\bf v}\vert{\bf v}') = \delta\left[ {\bf v}'-{\bf v}+ 2({\bf n}\cdot{\bf v})\,{\bf n}\right]
\ee
uniformly on the whole surface, which satisfies all the aforementioned properties.  In general, energy exchange happens during the lapse of time between the adsorption of the gas particles and their desorption.  If a local thermodynamic equilibrium at the temperature of the solid is assumed, the particles are expected to be desorbed with the corresponding Maxwell-Boltzmann distribution.  A very simple model achieving this condition is provided by the kernel
\be
p_{\bf r}({\bf v}\vert{\bf v}') = \vert{\bf n}\cdot{\bf v}\vert \; \frac{(m\beta_s)^2}{2\pi}\; \exp(-\beta_s\epsilon) \qquad \mbox{with}\qquad \epsilon = \frac{1}{2}m{\bf v}^2
\label{thermal-p}
\ee
which is independent of the other velocity ${\bf v}'$ and which also satisfies the required properties.
More realistic gas-surface kernels have been considered and discussed in the literature \cite{C00}. In principle, the form of the probability density should be determined by studying the quantum scattering process of the particles with the surface on the basis of their atomic structure \cite{KG86}.  

\vskip 0.5 cm

{\bf Remark.}  There exist further possible situations if surfaces at in relative motion one with respect to another, as in Couette-Taylor flows.  In such cases, the relative velocities of the surfaces are extra parameters controlling the nonequilibrium driving of the flow.  In some other circumstances, the linear momentum of the flow can be conserved, e.g., if the gas moves between parallel surfaces on which the particles undergo elastic collisions.  In this case, the velocity distributions of the gas incoming  from different reservoirs could have non-zero average velocities and an extra affinity is associated with the difference between the average velocities \cite{WVdBKL07}.  However, linear momentum is not conserved for typical geometries of the surfaces and gas-surface interactions so that the chemical affinity may be sufficient to characterize the nonequilibrium constraint on particle transport.

\section{The coarse-grained master equation}
\label{master_eq}

At the mesoscopic level of description, the quantities of interest are typically fluctuating.
This is for instance the case for transport processes or chemical reactions where the macroscopic equations of time evolution are replaced at the mesoscale by stochastic equations of Langevin type and their associated Fokker-Planck equation \cite{N72,vK81,G04,LL59,Z01,G05,OS06}.

Similar considerations have been applied to the Boltzmann equation and related kinetic equations \cite{N72,S49,BZ69,FU70,vK74,LK76,O78,BVdB80,EC81,vK81,G04}.  The idea consists in supposing that the fluctuating distribution function (\ref{f}) is ruled by a stochastic process.  In order to obtain the master equation of this process, the one-particle phase space of the position and velocity variables $({\bf r},{\bf v})$ can be partitioned into cells centered around the phase-space points $({\bf r}_{\alpha},{\bf v}_{\alpha})$.
These cells are of volume $\Delta r^3\Delta v^3$.
The number of particles in the cell $\alpha$ at the current time $t$ is defined by
\be
N_{\alpha}(t) \equiv \int_{\alpha} f({\bf r},{\bf v},t) \, d{\bf r} \, d{\bf v} \simeq f({\bf r}_{\alpha},{\bf v}_{\alpha},t) \, \Delta r^3 \Delta v^3
\ee
Since the distribution function $f$ is fluctuating, these particle numbers are random variables undergoing jumps every time a particle exits or enters the corresponding cell.

The probability to find $\{N_\alpha\}$ particles in the cells $\{\alpha\}$ at the current time $t$ is denoted as
\be
P = P({\bf N}) = P\left(\{ N_\alpha\}\right)
\label{P(N)}
\ee
When transitions happen, particles are exchanged between the cells so that it is useful to introduce the rising and lowering operators
\be
\hat E_{\alpha}^{\pm 1}\Phi(...,N_\alpha,...) = \Phi(...,N_\alpha\pm 1,...) 
\label{E}
\ee
acting on any function $\Phi({\bf N})=\Phi(\{N_\alpha\})$ by adding or removing one particle in the cell $\alpha$.  These operators can be written as
\be
\hat E_{\alpha}^{\pm 1} = \exp\left( \pm \frac{\partial}{\partial N_\alpha}\right)
\ee
which implies that they are adjoint to each other:
\be
\left( \hat E_\alpha^{\pm 1}\right)^{\dagger} = \hat E_\alpha^{\mp 1}
\label{E-dag}
\ee

The master equation ruling the probability distribution (\ref{P(N)}) takes the following form \cite{vK74,BVdB80,vK81}
\be
\frac{dP}{dt} = \sum_{\lambda\rho} W_{\lambda\rho}^{\rm FRS}\left( \hat E_{\lambda}^{-1}\hat E_{\rho}^{+1}-1\right) N_{\rho} P + \sum_{\lambda\mu\rho\sigma} W_{\lambda\mu\rho\sigma}^{\rm C}\left( \hat E_{\lambda}^{-1}\hat E_{\mu}^{-1}\hat E_{\rho}^{+1}\hat E_{\sigma}^{+1}-1\right) N_{\rho} N_{\sigma} P
\label{master}
\ee
with the transition rates due to the free flow (F), the contacts with the reservoirs (R), and the interactions with the surfaces (S):
\be
W_{\lambda\rho}^{\rm FRS}=W_{\lambda\rho}^{\rm F}+W_{\lambda\rho}^{\rm R}+W_{\lambda\rho}^{\rm S}
\ee
and those due to the binary collisions $W_{\lambda\mu\rho\sigma}^{\rm C}$.

The time evolution of the average particle numbers
\be
\langle N_{\alpha}\rangle \equiv \sum_{\bf N} N_\alpha \, P({\bf N})
\ee
and their higher statistical moments is ruled by a hierarchy of equations that can be deduced from the  master equation~(\ref{master}) \cite{vK81}.    For dilute gases, the second moments can be supposed to be factorized in terms of the averages themselves, in which case these latter quantities obey the following closed mean-field equations:
\be
\frac{d}{dt} \langle N_{\alpha}\rangle = \sum_{\rho} W_{\alpha\rho}^{\rm FRS}\, \langle N_{\rho}\rangle
- \langle N_{\alpha}\rangle\, \sum_\lambda W_{\lambda\alpha}^{\rm FRS} + 2 \sum_{\mu\rho\sigma} W_{\alpha\mu\rho\sigma}^{\rm C} \, \langle N_{\rho}\rangle\, \langle N_{\sigma}\rangle - 2 \, \langle N_{\alpha}\rangle\sum_{\lambda\mu\sigma} W_{\lambda\mu\alpha\sigma}^{\rm C} \, \langle N_{\sigma}\rangle
\label{MFE}
\ee
Although this mean-field kinetic equation is nonlinear because of the binary collisions, the master equation is linear and can be written as
\be
\frac{dP}{dt}= \hat L \, P
\ee
in terms of the linear operator $\hat L$ given by Eq.~(\ref{master}).  We notice that this operator is composed of two terms, $\hat L = \hat L^{\rm FRS}+\hat L^{\rm C}$, one corresponding to the motion of independent particles and the other to the binary collisions.  

At equilibrium, the stationary solutions of the coarse-grained master equation (\ref{master}) are given by multiple Poisson distributions of the following form:
\be
P_{\rm eq}({\bf N}) = \prod_\alpha {\rm e}^{-\langle N_\alpha\rangle} \frac{\langle N_\alpha\rangle^{N_\alpha}}{N_\alpha !}\qquad\mbox{with}\qquad \langle N_\alpha\rangle = f_{\rm eq}({\bf v}_\alpha)\, \Delta r^3 \, \Delta v^3
\label{P_eq}
\ee
being the average values of the particle numbers in every cell $\alpha$ given in terms of the overall equilibrium Maxwell-Boltzmann distribution (\ref{MB_eq}) at the uniform temperature $T=(k_{\rm B}\beta)^{-1}$ and density $n$.  This multiple Poisson distribution represents the equilibrium state because it corresponds to the Maxwell-Boltzmann distribution (\ref{MB_eq}) that is the equilibrium stationary solution of Boltzmann's equation (\ref{B_eq}).  

\subsection{In the bulk of the flow}

The transition rate of the events during which one particle is flowing in free flight $\rho\to\lambda$ from the cell $\rho$ to the cell $\lambda$ is given by
\be
W_{\lambda\rho}^{\rm F} = \frac{1}{\Delta r} \, ({\bf n}_{\lambda\rho}\cdot{\bf v}) \, \theta({\bf n}_{\lambda\rho}\cdot{\bf v}) \, \delta_{{\bf r}_\lambda,{\bf r}_\rho+\Delta r \, {\bf n}_{\lambda\rho}} \, \delta_{{\bf v}_\lambda,{\bf v}_\rho} 
\label{W_F}
\ee
where ${\bf n}_{\lambda\rho}$ is a unit vector directed from the center ${\bf r}_\rho$ of the cell $\rho$ to the center ${\bf r}_\lambda$ of the cell $\lambda$ \cite{BVdB80,G04}.  The Kronecker delta $\delta_{{\bf r}_\rho,{\bf r}_\lambda-\Delta r \, {\bf n}_{\lambda\rho}}$ expresses the fact that the transition occurs between next-neighboring cells such that ${\bf r}_\lambda={\bf r}_\rho+\Delta r \, {\bf n}_{\lambda\rho}$.  The other Kronecker delta $\delta_{{\bf v}_\rho,{\bf v}_\lambda}$ means that the velocity remains constant during the free flight. In the case where the particle undergoes an acceleration, $\delta_{{\bf v}_\rho,{\bf v}_\lambda}$ should be replaced by the appropriate expression taking into account the force field ${\bf F}({\bf r})$ acting on the particle.  Since the one-particle motion is time-reversal symmetric, the corresponding rates have the symmetry:
\be
\mbox{time-reversal symmetry:}\qquad W_{\lambda\rho}^{\rm F}=W_{\rho^{\rm T}\lambda^{\rm T}}^{\rm F} \label{T-sym-WF}\\
\ee
where $\alpha^{\rm T}=({\bf r}_\alpha,{\bf v}_\alpha)^{\rm T}=({\bf r}_\alpha,-{\bf v}_\alpha)$ is the time-reversed cell corresponding to the cell $\alpha$.  The symmetry (\ref{T-sym-WF}) holds because ${\bf n}_{\lambda\rho}=-{\bf n}_{\rho\lambda}$.  Moreover, the rates (\ref{W_F}) satisfy the property of reciprocity:
\be
W_{\lambda\rho}^{\rm F} \; \langle N_\rho\rangle_{\rm eq} = W_{\rho^{\rm T}\lambda^{\rm T}}^{\rm F} \; \langle N_\lambda\rangle_{\rm eq} 
\label{reciprocity-WF}
\ee
where $\langle N_\alpha\rangle_{\rm eq}$ denotes the average number of particles in the cell $\alpha$ at the equilibrium state (\ref{P_eq}).

On the other hand, the rate of the transition $\rho\sigma\to\lambda\mu$ due to binary collisions is given by\be
W_{\lambda\mu\rho\sigma}^{\rm C} = \frac{1}{2\Delta r^3\Delta v^6} \, \int_\lambda d{\bf v}_1\int_\mu d{\bf v}_2 \int_\rho d{\bf v}_1' \int_\sigma d{\bf v}_2' \; w\left({\bf v}_1,{\bf v}_2\vert{\bf v}_1',{\bf v}_2'\right)
\, \delta_{{\bf r}_\mu,{\bf r}_\lambda} \, \delta_{{\bf r}_\rho,{\bf r}_\lambda}\, \delta_{{\bf r}_\sigma,{\bf r}_\lambda} 
\label{W_C}
\ee
These transitions do not change the positions but modify the velocities according to the collision rule.
We notice that the symmetries (\ref{T-sym})-(\ref{PT-sym}) of the transition rates of Boltzmann's equation imply similar symmetries for the rates (\ref{W_C}):
\bea
\mbox{time-reversal symmetry:}\qquad && W_{\lambda\mu\rho\sigma}^{\rm C}=W_{\rho^{\rm T}\sigma^{\rm T}\lambda^{\rm T}\mu^{\rm T}}^{\rm C} \label{T-sym-WC}\\
\mbox{inversion symmetry:}\qquad && W_{\lambda\mu\rho\sigma}^{\rm C}=W_{\rho\sigma\lambda\mu}^{\rm C} \label{PT-sym-WC}
\eea
Moreover, since the binary-collision terms in the master equation (\ref{master}) are summed over the four indices, the following symmetries also hold \cite{LK76}
\be
W_{\lambda\mu\rho\sigma}^{\rm C}=W_{\mu\lambda\rho\sigma}^{\rm C}=W_{\lambda\mu\sigma\rho}^{\rm C}=W_{\mu\lambda\sigma\rho}^{\rm C}
\ee
The rates (\ref{W_C}) also satisfy the property of reciprocity:
\be
W_{\lambda\mu\rho\sigma}^{\rm C} \; \langle N_\rho\rangle_{\rm eq}  \; \langle N_\sigma\rangle_{\rm eq} = W_{\rho^{\rm T}\sigma^{\rm T}\lambda^{\rm T}\mu^{\rm T}}^{\rm C} \; \langle N_\lambda\rangle_{\rm eq} \; \langle N_\mu\rangle_{\rm eq} 
\label{reciprocity-WC}
\ee
so that the coarse-grained master equation (\ref{master}) preserves the local thermodynamic equilibrium.
The multiple Poisson distribution (\ref{P_eq}) is symmetric under time reversal because the Maxwell-Boltzmann distribution is invariant under velocity reversal ${\bf v}\to -{\bf v}$ so that $\langle N_\alpha\rangle=\langle N_{\alpha^{\rm T}}\rangle$ at equilibrium.  Therefore, the principle of detailed balancing holds, according to which the probabilities of every transition are equal to those of the opposite transition at the thermodynamic equilibrium \cite{vK81,G04,vK57}.  In the limit $\Delta r^3\Delta v^3\to 0$ where the volume of the cells vanishes, we recover Boltzmann's equation (\ref{B_eq}) in the bulk of the flow from Eq.~(\ref{MFE}) with the transition rates (\ref{W_F}) and (\ref{W_C}).

\subsection{At the boundaries}

\subsubsection{At the reservoirs}

\begin{figure}[h]
\begin{center}
\includegraphics[scale=0.45]{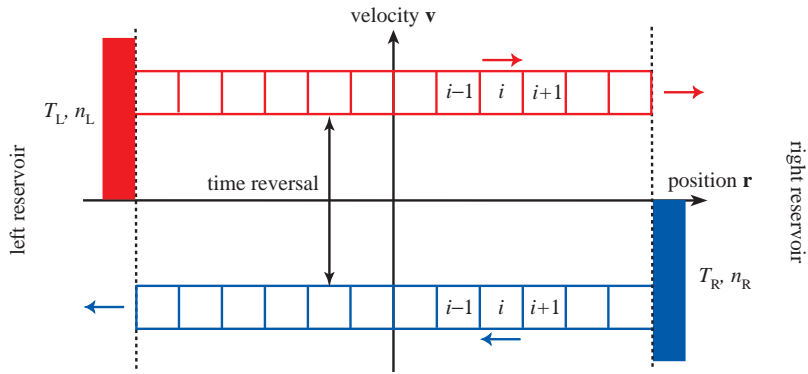}
\caption{Schematic phase portrait of the free motion between two reservoirs from which the particles are incoming with different temperatures and densities.}
\label{fig2}
\end{center}
\end{figure}

For open systems, the partition into cells $\{\alpha\}$ should be carried out in a finite volume between the reservoirs.  The particles are entering and leaving the system from and to the reservoirs.  These specific motions are ruled by the flow operator $\hat L^{\rm FRS}$.  The key point is that this operator depends on the boundary conditions imposed to the system by the temperatures and particle densities of the flows incoming from the reservoirs.  We first consider the simplest case of free flight between two reservoirs, as schematically depicted in Fig.~\ref{fig2}. 

The time-reversal transformation $({\bf r},{\bf v})\to({\bf r},-{\bf v})$ maps the trajectories with positive velocity coming from the left-hand reservoir onto those with negative velocity coming from the right-hand reservoir.  The phase space between both reservoirs is partitioned into cells.  The cells at the boundaries exchange particles with the reservoirs.  The particles are dropped into the corresponding reservoir where they exit the open system.  However, they enter into the system with a Maxwell-Boltzmann distribution at the temperature and density of the reservoir from which they come.  Accordingly, the rate of the transition $N\to N+1$ in the cell at the boundary of an incoming flow is determined in terms of the temperature and density of the nearby reservoir in analogy with the master-equation formulation of diffusion \cite{G05}.  In the simple geometry of Fig.~\ref{fig2}, the part of the flow operator for particles in free flight at the velocities $\pm v$ acts on functions $\Phi(\{N_{i,\pm}\}_{i=1}^I)$ where $N_{i,\pm}$ is the number of particles in the cell at the position $i$ and velocity $\pm v$.  The free flight from one reservoir to the other is partitioned into $I$ cells of size $\Delta r$ so that the cell index $i$ runs from $i=1$ to $i=I$.  For the particles of velocity $\pm v$, the flow operator with the inlets from the reservoirs would be:
\bea
\hat L^{\rm FR}\Phi &=& \frac{v}{\Delta r} \left[ (\hat E_{1,+}^{-1}-1)\langle N_{+}\rangle_{\rm L} \Phi +
\sum_{i=1}^{I-1} (\hat E_{i,+}^{+1}\hat E_{i+1,+}^{-1}-1) N_{i,+} \Phi + (\hat E_{I,+}^{+1}-1) N_{I,+} \Phi \right. \nonumber\\
&& \qquad\qquad +\left. (\hat E_{1,-}^{+1}-1)N_{1,-} \Phi +
\sum_{i=1}^{I-1} (\hat E_{i+1,-}^{+1}\hat E_{i,-}^{-1}-1) N_{i+1,-} \Phi + (\hat E_{I,-}^{-1}-1) \langle N_{-}\rangle_{\rm R} \Phi \right] \label{flow_op}
\eea
where $\langle N_{+}\rangle_{\rm L}=f_{\rm L}\Delta r^3\Delta v^3$ is the average number of particles incoming with velocity $+v$ from the left-hand reservoir at the temperature $T_{\rm L}$ and density $n_{\rm L}$ while $\langle N_{-}\rangle_{\rm R}=f_{\rm R}\Delta r^3\Delta v^3$ is the average number of particles incoming with velocity $-v$ from the right-hand reservoir at the temperature $T_{\rm R}$ and density $n_{\rm R}$.  These average numbers can be expressed in terms of the equilibrium Maxwell-Boltzmann distribution function (\ref{MB_distrib}) in the corresponding reservoir $s={\rm L},{\rm R}$.  For simplicity, only the particles moving at the kinetic energy $\epsilon=m v^2/2$ in one direction are here considered.

The average particle numbers at the reservoirs are related to the thermal and chemical affinities (\ref{A_E}) and (\ref{A_N}) according to
\be
A = \epsilon \, A_E + A_N = \ln \frac{\langle N_{+}\rangle_{\rm L}}{\langle N_{-}\rangle_{\rm R}} = \ln\frac{f_{\rm L}}{f_{\rm R}}
\label{A}
\ee
The stationary solution of the master equation (\ref{master}) for non-vanishing values of the affinities gives the invariant probability distribution of the corresponding nonequilibrium steady state.

At equilibrium, the stationary solution of the master equation (\ref{master}) is the multiple Poisson distribution (\ref{P_eq}) so that the temperatures and chemical potentials are equal and all the affinities are vanishing.

\subsubsection{At the surfaces}

In the case of elastic collisions on a surface, the particle follows a deterministic orbit which is piecewise straight. Let us consider an orbit coming from the right-hand reservoir, colliding on the surface and going back to the same reservoir.  The speed $\Vert{\bf v}\Vert$ is conserved during the whole flight although the velocity direction undergoes the specular reflection given by Eq.~(\ref{specular}).  This is the case for every orbit of this kind.  Accordingly, the motion along such orbits can be described as before by a partition into $I$ cells of size $\Delta r$ so that the cell index $i$ runs from $i=1$ to $i=I$.  For an orbit interrupted by a specular reflection, the flow operator would be:
\bea
\hat L^{\rm FS}\Phi &=& \frac{v}{\Delta r} \left[ (\hat E_{1,-}^{+1}\hat E_{1,+}^{-1}-1) N_{1,-} \Phi +
\sum_{i=1}^{I-1} (\hat E_{i,+}^{+1}\hat E_{i+1,+}^{-1}-1) N_{i,+} \Phi + (\hat E_{I,+}^{+1}-1) N_{I,+} \Phi \right. \nonumber\\
&& \qquad\qquad \left.  +
\sum_{i=1}^{I-1} (\hat E_{i+1,-}^{+1}\hat E_{i,-}^{-1}-1) N_{i+1,-} \Phi + (\hat E_{I,-}^{-1}-1) \langle N_{-}\rangle_{\rm R} \Phi \right] \label{flow_op_specular}
\eea
Comparing with Eq.~(\ref{flow_op}), we see that all the terms are now of propagative type except the terms at the right-hand reservoirs, but they are the same as before.   This means that the operator is of the same type as for a free flight from the right-hand reservoir back to itself and no difference in temperature or particle density manifests itself in this case.

The situation is different in the case where gas-surface interactions induce energy exchange described for instance by Eq.~(\ref{term-S}) with the simple thermal kernel (\ref{thermal-p}).  To fix the ideas, the surface is taken as $z=0$ so that the normal unit vector is ${\bf n}=(0,0,1)$.  If we discretize position and velocity, the corresponding operator in the coarse-grained master equation is given by
\be
\hat L^{\rm S}\Phi = \sum_{\bf r} \delta_{z,0}  \sum_{\scriptstyle{\bf v} \atop \scriptstyle v_z>0}  \sum_{\scriptstyle{\bf v}' \atop \scriptstyle v'_z<0} \Delta v^3  \, \frac{\vert v'_z\vert}{\Delta r} \; p({\bf v}\vert{\bf v}') \, \left( \hat E_{{\bf r}{\bf v}'}^{+1} \, \hat E_{{\bf r}{\bf v}}^{-1} - 1 \right) N_{{\bf r}{\bf v}} \, \Phi
\label{S_op}
\ee
expressing the adsorption of a particle of velocity ${\bf v}'$ with $v'_z<0$ and its desorption at the velocity $\bf v$ with $v_z>0$ distributed according to a kernel satisfying the three properties (\ref{p-norm}), (\ref{p-eq}), and (\ref{p-reciprocity}) inducing a thermalization at the temperature $T_s$ of the surface.  This operator has to be added to the free-flight operator $\hat L^{\rm F}$.

\section{The fluctuating Boltzmann equation}
\label{fluctuating_Boltzmann}

In the diffusive limit, a functional master equation of Fokker-Planck type is deduced from the coarse-grained  master equation (\ref{master}).  The diffusive limit is reached by considering cells containing sufficiently large numbers of particles that they can be described by Gaussian distributions while taking the limit of arbitrarily small cells to reach the continuum limit.  The resulting functional equation of Fokker-Planck type rules the time evolution for the probability density functional that the random distribution function (\ref{f}) takes some given form at the current time $t$: ${\cal P}[f]={\cal P}[f({\bf r},{\bf v})]$. In the basic situation without contact with reservoirs or thermalizing surfaces, the functional master equation writes:
\be
\frac{\partial}{\partial t} {\cal P}[f] = - \int d{\bf r}\, d{\bf v} \, \frac{\delta}{\delta f({\bf r},{\bf v})} \, {\cal A}[f]\, {\cal P}[f] + \frac{1}{2} \int d{\bf r}\, d{\bf v}\, d{\bf r}'\, d{\bf v}' \, \frac{\delta^2}{\delta f({\bf r},{\bf v})\,\delta f({\bf r}',{\bf v}')} \, {\cal B}[f]\, {\cal P}[f] 
\label{FME}
\ee
with the deterministic kinetic rate
\be
{\cal A}[f] = - {\bf v}\cdot\frac{\partial f}{\partial{\bf r}} 
- \frac{\bf F}{m}\cdot\frac{\partial f}{\partial{\bf v}} + \int d{\bf v}_2 \; d{\bf v}_1' \; d{\bf v}_2' 
\; w\left({\bf v}_1,{\bf v}_2\vert{\bf v}_1',{\bf v}_2'\right) \; \left( f_1'\, f_2'- f_1\, f_2\right)
\label{A[f]}
\ee
and the diffusivity coefficient
\bea
{\cal B}[f] &=& \delta({\bf r}-{\bf r}') \int d{\bf v}_2 \; d{\bf v}_3 \; d{\bf v}_4 \; w\left({\bf v}_1,{\bf v}_2\vert{\bf v}_3,{\bf v}_4\right) \; \left( f_1\, f_2+ f_3\, f_4\right) \nonumber\\
&&\qquad\qquad\qquad\times
\left[ \delta({\bf v}_1-{\bf v}') +\delta({\bf v}_2-{\bf v}') - \delta({\bf v}_3-{\bf v}') - \delta({\bf v}_4-{\bf v}')\right] 
\label{B[f]}
\eea
and where $\delta/\delta f$ denotes a functional derivative \cite{BVdB80,vK81,G04}.
We notice that this functional master equation is of diffusion type as for the usual Fokker-Planck equations associated to Langevin stochastic differential equations. The average distribution function (\ref{<f>}) obeys an equation at the top of a hierarchy of moments equations.  If the noise amplitude is low enough, this hierarchy may be truncated and Boltzmann's equation (\ref{B_eq}) is thus recovered. To go beyond this mean-field approximation and consider the effects of the fluctuations, the functional master equation (\ref{FME}) should be used. 

If the system is open, the injection of particles from the reservoirs is described at the mean-field level by the term (\ref{B_eq_R}) in the Boltzmann equation (\ref{B_eq}).  For the coarse-grained master equation (\ref{master}), this is taken into account by the boundary terms in the flow operator (\ref{flow_op}).  In the diffusive limit, the following extra terms must be added to the deterministic kinetic rate (\ref{A[f]}) and the diffusivity coefficient (\ref{B[f]}):
\bea
&& {\cal A}^{\rm R}[f] = \sum_s \int_s d^2S \; \delta({\bf r}-{\bf r}_s) \; ({\bf n}\cdot{\bf v}) \; \theta({\bf n}\cdot{\bf v}) \; \left[ f_s({\bf v})- f({\bf r},{\bf v})\right] \\
&& {\cal B}^{\rm R}[f] = \delta({\bf r}-{\bf r}')\, \delta({\bf v}-{\bf v}')\, \sum_s \int_s d^2S \; \delta({\bf r}-{\bf r}_s) \; ({\bf n}\cdot{\bf v}) \; \theta({\bf n}\cdot{\bf v}) \; \left[ f_s({\bf v})+ f({\bf r},{\bf v})\right]
\eea
Similarly, the presence of a thermalizing surface also contributes by corresponding terms ${\cal A}^{\rm S}[f]$ and ${\cal B}^{\rm S}[f]$.

The analogue of the Langevin equation associated with the functional Fokker-Planck equation (\ref{FME}) is the following stochastic integro-differential equation of It\^o type for the random distribution function (\ref{f}):
\be
\frac{\partial f}{\partial t} + {\bf v}\cdot\frac{\partial f}{\partial{\bf r}} 
+ \frac{\bf F}{m}\cdot\frac{\partial f}{\partial{\bf v}} = \int d{\bf v}_2 \; d{\bf v}_1' \; d{\bf v}_2' 
\; w\left({\bf v}_1,{\bf v}_2\vert{\bf v}_1',{\bf v}_2'\right) \; \left( f_1'\, f_2'- f_1\, f_2\right) + g({\bf r},{\bf v},t)
\label{F_B_eq}
\ee
where $g({\bf r},{\bf v},t)$ is a Gaussian white noise satisfying:
\bea
&&\langle g({\bf r},{\bf v},t)\rangle = 0 \label{aver_g}\\
&&\langle g({\bf r},{\bf v},t) \, g({\bf r}',{\bf v}',t')\rangle 
=\delta(t-t')\; \delta({\bf r}-{\bf r}') \int d{\bf v}_2 \; d{\bf v}_3 \; d{\bf v}_4 \; w\left({\bf v}_1,{\bf v}_2\vert{\bf v}_3,{\bf v}_4\right) \; \left( f_1\, f_2+ f_3\, f_4\right) \nonumber\\
&&\qquad\qquad\qquad\qquad\qquad\qquad\qquad\qquad\qquad\times
\left[ \delta({\bf v}_1-{\bf v}') +\delta({\bf v}_2-{\bf v}') - \delta({\bf v}_3-{\bf v}') - \delta({\bf v}_4-{\bf v}')\right] \label{var_g}
\eea
as shown explicitly in Ref.~\cite{BVdB80}.
In an open system, the contact with the reservoirs and thermalizing surfaces adds the contributions
\be
\langle g({\bf r},{\bf v},t) \, g({\bf r}',{\bf v}',t')\rangle^{\rm R,S}=\delta(t-t')\; {\cal B}^{\rm R,S}[f]
\ee
to the variance (\ref{var_g}) of the noise.

Since the functional master equation (\ref{FME}) can be viewed as the continuum limit of the coarse-grained master equation (\ref{master}), we shall use this latter which is more convenient to handle mathematically.

\section{The time-reversal symmetry relation and its consequences}
\label{TRsym}

\subsection{The theorem}

To fix the framework, we consider the open system with two reservoirs sketched in Fig.~\ref{fig1}a in a stationary state of affinities ${\bf A}=\{A_E,A_N\}$ given by Eqs.~(\ref{A_E}) and (\ref{A_N}).  The transport of energy and particles between the reservoirs is characterized in terms of the instantaneous currents:
\bea
j_E(t) &=& \sum_{k=-\infty}^{+\infty} \sigma_k \, \epsilon_k \, \delta(t-t_k)  \label{J_E}\\
j_N(t) &=& \sum_{k=-\infty}^{+\infty} \sigma_k \, \delta(t-t_k)  \label{J_N}
\eea
where $\{ t_k\}_{k=-\infty}^{+\infty}$ is the sequence of random times when a particle enters or exits the open system at the left-hand reservoir with its energy $\epsilon_k$.  The sign of the exchange is taken into account by setting $\sigma=+1$ if the particle enters and $\sigma=-1$ if the particle exits.

We introduce the cumulant generating function of the fluctuating currents as
\be
Q_{\bf A}(\pmb{\lambda}) \equiv \lim_{t\to\infty} \, -\frac{1}{t} \ln \left\langle \exp \left[ - \pmb{\lambda}\cdot\int_0^t {\bf j}(\tau)\, d\tau \right]\right\rangle_{\bf A}
\label{Q}
\ee
with the counting parameters $\pmb{\lambda}=\{\lambda_E,\lambda_N\}$, the fluctuating instantaneous currents ${\bf j}(t)=\{j_E(t),j_N(t)\}$, and where $\langle\cdot\rangle_{\bf A}$ denotes the statistical average in the stationary state of affinities $\bf A$.
This generating function can be obtained as solution of the eigenvalue problem:
\be
\hat L_{\pmb{\lambda}} \, \Psi = - Q_{\bf A}(\pmb{\lambda}) \, \Psi
\label{eigen}
\ee
for the modified master-equation operator:
\be
\hat L_{\pmb{\lambda}} \equiv {\rm e}^{-\pmb{\lambda}\cdot {\bf G}} \, \hat L \, {\rm e}^{\pmb{\lambda}\cdot {\bf G}}
\label{modif-L}
\ee
where ${\bf G}=\int_0^t {\bf j}(\tau)\, d\tau$ are observables called Helfand moments in transport theory and such that their time derivative is equal to the instantaneous current ${\bf j}(t)=d{\bf G}/dt$ \cite{GD95}.  We have the following:

\vskip 0.5 cm

{\bf Theorem.} {\it The modified operator Eq.~(\ref{modif-L}) corresponding to the master equation (\ref{master}) obeys the symmetry relation:}
\be
\eta^{-1} \Theta \, \hat L_{\pmb{\lambda}} \left( \Theta \, \eta \, \Phi\right) = \hat L_{{\bf A}-\pmb{\lambda}}^{\dagger}\Phi
\label{sym-rel}
\ee
{\it where $\Phi({\bf N})$ denotes some arbitrary function, $\Theta$ is the involution performing the time-reversal transformation:}
\be
\Theta \, \Phi(\{N_{\alpha}\}) = \Phi(\{N_{\alpha^{\rm T}}\}) \qquad\mbox{\it so that}\qquad \Theta^2 =1
\label{Theta}
\ee
{\it and $\eta({\bf N})$ is the multiple Poisson distribution:}
\be
\eta({\bf N}) = \prod_\alpha {\rm e}^{-\langle N_\alpha\rangle} \frac{\langle N_\alpha\rangle^{N_\alpha}}{N_\alpha !} \qquad\mbox{\it with}\qquad
\langle N_\alpha\rangle = f_{\rm R}({\bf v}_\alpha)\, \Delta r^3 \, \Delta v^3
\label{eta}
\ee
{\it given in terms of the Maxwell-Boltzmann distribution $f_{\rm R}({\bf v})$ of some reference reservoir, for instance the right-hand reservoir at the temperature $T_{\rm R}$ and density $n_{\rm R}$.}

\subsection{Proof of the theorem}

\subsubsection{Open system in contact with two reservoirs}

This theorem is established as follows.  The first step is to notice that only the flow part of the master-equation operator is modified by Eq.~(\ref{modif-L}):
\be
\hat L_{\lambda_E,\lambda_N} = \hat L_{\lambda_E,\lambda_N}^{\rm FR}+\hat L^{\rm C}
\ee
Indeed, the operator $\hat L^{\rm C}$ due to the binary collisions acts locally exchanging particles among all the cells corresponding to the same position.  Since the observables (\ref{J_E}) and (\ref{J_N}) are non zero for particle exchanges at the boundary with the left-hand reservoir, there is no modification of the binary-collision operator.  

Now, the binary-collision operator always obeys the symmetry relation:
\be
\eta^{-1} \Theta \, \hat L^{\rm C} \left( \Theta \, \eta \, \Phi\right) = \hat L^{{\rm C}\dagger}\Phi
\label{sym-rel-C}
\ee
as shown in Appendix \ref{AppA} by using the property that the lowering and rising operators (\ref{E}) act as follows on the product of the multiple Poisson distribution (\ref{eta}) with an arbitrary function $\Phi$:
\bea
&&\hat E_\alpha^{-1} \, \eta \, \Phi = \frac{N_\alpha}{\langle N_\alpha\rangle} \; \eta \; \hat E_\alpha^{-1} \,\Phi \label{E--eta}\\
&&\hat E_\alpha^{+1}\, \eta \, \Phi = \frac{\langle N_\alpha\rangle}{N_\alpha+1} \; \eta \; \hat E_\alpha^{+1} \,\Phi 
\label{E+-eta}
\eea

On the other hand, the flow operator (\ref{flow_op}) is modified by the transformation (\ref{modif-L}) because it contributes to the exchange of particles with the left-hand reservoir where the currents (\ref{J_E}) and (\ref{J_N}) are defined. For particles at the energy $\epsilon=mv^2/2$, the flow operator (\ref{flow_op}) is modified into
\bea
\hat L_{\lambda_E,\lambda_N}^{\rm FR}\Phi &=& \frac{v}{\Delta r} \left[ ({\rm e}^{-\lambda}\,\hat E_{1,+}^{-1}-1)\langle N_{+}\rangle_{\rm L} \Phi + \sum_{i=1}^{I-1} (\hat E_{i,+}^{+1}\hat E_{i+1,+}^{-1}-1) N_{i,+} \Phi 
 + (\hat E_{I,+}^{+1}-1) N_{I,+} \Phi \right. \nonumber\\
&& \qquad\qquad +\left. ({\rm e}^{+\lambda}\,\hat E_{1,-}^{+1}-1)N_{1,-} \Phi +
\sum_{i=1}^{I-1} (\hat E_{i+1,-}^{+1}\hat E_{i,-}^{-1}-1) N_{i+1,-} \Phi + (\hat E_{I,-}^{-1}-1) \langle N_{-}\rangle_{\rm R} \Phi \right]
\eea
with the counting parameter $\lambda=\epsilon \lambda_E+\lambda_N$.  As aforementioned, the multiple Poisson distribution (\ref{eta}) is symmetric under time reversal, $\Theta\eta=\eta$, where the time-reversal transformation acts as $\Theta N_{i,\pm}=N_{i,\mp}$, exchanging the velocities $\pm v$ at every position $i=1,2,...,I$.  Moreover, we have the properties (\ref{E-dag}), (\ref{E--eta}), and (\ref{E+-eta}) for the lowering and rising operators.  Therefore, the transformation in the left-hand side of Eq.~(\ref{sym-rel}) has the effect of attributing the terms of positive velocity to those of negative velocity and vice versa. Furthermore, the boundary term with ${\rm e}^{-\lambda}\langle N_{+}\rangle_{\rm L}$ is transformed into ${\rm e}^{-\lambda}\langle N_{+}\rangle_{\rm L}/\langle N_{1,-}\rangle_{\rm R}$ and the boundary term with ${\rm e}^{+\lambda}$ into ${\rm e}^{+\lambda}\langle N_{1,+}\rangle_{\rm R}$.  Now, the left-hand reservoir is at the temperature $T_{\rm L}$ and density $n_{\rm L}$ while the right-hand reservoir is used as reference so that $\langle N_{i,\pm}\rangle_{\rm R}=f_{\rm R}\Delta r^3\Delta v^3$ for all $i=1,2,...,I$ according to Eq.~(\ref{eta}), whereupon we have the equalities:
\bea
{\rm e}^{-\lambda}\, \frac{\langle N_{+}\rangle_{\rm L}}{\langle N_{1,-}\rangle_{\rm R}} &=& {\rm e}^{A-\lambda}\\
{\rm e}^{+\lambda}\langle N_{1,+}\rangle_{\rm R} &=& {\rm e}^{-(A-\lambda)}\, \langle N_{+}\rangle_{\rm L}
\eea
with the affinity $A$ defined by Eq.~(\ref{A}).  Consequently, the two boundary terms with the counting parameter $\lambda$ may exchange their role if we carry out the transformation $\lambda\to A-\lambda$. 
We thus obtain the symmetry relation of the flow operator:
\be
\eta^{-1} \Theta \, \hat L_{\lambda_E,\lambda_N}^{\rm FR} \left( \Theta \, \eta \, \Phi\right) = \hat L_{A_E-\lambda_E,A_N-\lambda_N}^{{\rm FR}\dagger}\Phi
\label{sym-rel-F}
\ee
Combining the results (\ref{sym-rel-C}) and (\ref{sym-rel-F}), the symmetry relation (\ref{sym-rel}) is finally proved.

\subsubsection{Open system also in contact with a thermalizing surface}

Here, we consider the case where the system is open and includes a thermalizing surface described by the operator (\ref{S_op}).  Since energy exchange may occur at the surface, we have to introduce an extra counting parameter $\lambda_{E}^{\rm S}$ corresponding to an instantaneous current similar to Eq.~(\ref{J_E}) but defined at the surface.  Accordingly, the operator is modified to
\be
\hat L^{\rm S}_{\lambda_{E}^{\rm S}}\Phi = \sum_{\bf r} \delta_{z,0}  \sum_{\scriptstyle{\bf v} \atop \scriptstyle v_z>0}  \sum_{\scriptstyle{\bf v}' \atop \scriptstyle v'_z<0} \Delta v^3  \, \frac{\vert v'_z\vert}{\Delta r} \; p({\bf v}\vert{\bf v}') \, \left[{\rm e}^{\lambda_{E}^{\rm S}(\epsilon'-\epsilon)}\hat E_{{\bf r}{\bf v}'}^{+1} \, \hat E_{{\bf r}{\bf v}}^{-1} - 1 \right] N_{{\bf r}{\bf v}} \, \Phi
\label{S_op_modif}
\ee
where $\epsilon=m{\bf v}^2/2$ and $\epsilon'=m{\bf v}'^2/2$.  The identity
\be
\eta^{-1} \Theta \, \hat L_{\lambda_{E}^{\rm S}}^{\rm S} \left( \Theta \, \eta \, \Phi\right) = \hat L_{A_E^{\rm S}-\lambda_{E}^{\rm S}}^{{\rm S}\dagger}\Phi
\label{sym-rel-S}
\ee
is obtained provided that the gas-surface interaction kernel satisfies:
\be
\frac{\vert v'_z\vert \, p({\bf v}\vert{\bf v}')}{\vert v_z\vert \, p(-{\bf v}'\vert -{\bf v})} = {\rm e}^{(A_E^{\rm S}-\beta_{\rm R})(\epsilon-\epsilon')}
\ee
This is always guaranteed thanks to the property of reciprocity (\ref{p-reciprocity}) if the affinity associated with the thermalizing surface is defined by
\be
A_E^{\rm S} = \beta_{\rm R}-\beta_s
\label{A_E^S}
\ee
where $\beta_s=(k_{\rm B}T_s)^{-1}$ characterizes the Maxwell-Boltzmann distribution (\ref{MB_distrib}) at the surface temperature $T_s$.

For an open system in contact with two reservoirs and a thermalizing surface, we have thus proved the symmetry relation (\ref{sym-rel}) with the counting parameters $\pmb{\lambda}=\{\lambda_E,\lambda_N,\lambda_{E}^{\rm S}\}$ and the affinities ${\bf A}=\{A_E,A_N,A_E^{\rm S}\}$ given by Eqs.~(\ref{A_E}), (\ref{A_N}), and (\ref{A_E^S}).

Such symmetry relations extend from the coarse-grained master equation to the fluctuating Boltzmann equation as long as this latter is the diffusive limit of the former together with their respective modified evolution operators.

\subsection{The current fluctuation theorem}

The consequence of the theorem is that the real eigenvalues of Eq.~(\ref{eigen}) and, in particular, the leading eigenvalue giving the generating function (\ref{Q}) also have the symmetry of the modified operator so that we get
the time-reversal symmetry relation for the cumulant generating function:
\be
Q_{\bf A}(\pmb{\lambda}) = Q_{\bf A}({\bf A}-\pmb{\lambda})
\label{FTC_Q}
\ee

If we denote by $P_{\bf A}({\bf J},t)$ the time-dependent probability that the instantaneous currents averaged over a time interval $t$ take the values 
\be
{\bf J}=\frac{1}{t} \, \int_0^t {\bf j}(\tau) \, d\tau
\ee
the fluctuation theorem for the currents is obtained:
\be
\frac{P_{\bf A}({\bf J},t)}{P_{\bf A}(-{\bf J},t)} \simeq_{t\to\infty} \exp \left( {\bf A}\cdot{\bf J} \, t\right) 
\label{FTC}
\ee
in the long-time limit for the fluctuating Boltzmann equation.  The equivalence between Eqs.~(\ref{FTC_Q}) and (\ref{FTC}) is demonstrated by noting that the cumulant generating function can be expressed in terms of the probability of the fluctuating currents as
\be
Q_{\bf A}(\pmb{\lambda}) \equiv \lim_{t\to\infty} \, -\frac{1}{t} \ln \int P_{\bf A}({\bf J},t) \, \exp(- \pmb{\lambda}\cdot{\bf J }\, t ) \; d{\bf J}
\label{Q-P}
\ee
so that Eq.~(\ref{FTC}) yields Eq.~(\ref{FTC_Q}) and vice versa.

As a consequence of the current fluctuation theorem, the generalizations of Green-Kubo formulae and Onsager reciprocity relations obtained in Ref.~\cite{AG07JSM} for the nonlinear response coefficients extend to the nonequilibrium flows of dilute and rarefied gases ruled by the fluctuating Boltzmann equation.

\section{Conclusions}
\label{conclusions}

In this paper, a time-reversal symmetry relation has been established for all the random currents of particle and energy carried by the particles of dilute or rarefied gases flowing in open systems ruled by the fluctuating Boltzmann equation.  For this purpose, the fluctuating Boltzmann equation is obtained as the Langevin equation corresponding to a coarse-grained master equation for the random numbers of particles in cells partitioning the one-particle phase space, in the limit of arbitrarily small cells.  The coarse-grained master equation has been obtained not only in the bulk of the fluid but also at its boundaries where the nonequilibrium constraints are imposed.  These constraints are characterized in terms of the thermodynamic forces or affinities defined from the differences of temperatures or particle densities between gas reservoirs or, possibly, thermalizing surfaces.  

The time-evolution operator of the coarse-grained master equation is composed of several terms corresponding to the different processes taking place in the bulk or at the boundaries.  This linear operator is modified by introducing the observables counting the transfers of particles and energy at the boundaries and the associated counting parameters.  The cumulant generating function of the fluctuating currents is given as the leading eigenvalue of this modified operator.  The central result is obtained by showing that the modified operator obeys the symmetry relation (\ref{sym-rel}) involving the counting parameters and the affinities.  As a consequence, the cumulant generating function has the symmetry (\ref{FTC_Q}) and the fluctuation theorem (\ref{FTC}) is thus established for all the currents.  These relations find their origin in the time-reversal symmetry of the underlying microscopic dynamics.

The fluctuation relation (\ref{FTC}) implies the non-negativity of the entropy production defined as the sum of the affinities multiplied by the average values of the currents flowing across the open system.  Furthermore, the fluctuation relation also implies the Green-Kubo formulae and Onsager reciprocity relations as well as their generalizations beyond linear response theory \cite{AG07JSM}, which thus hold for dilute or rarefied gases ruled by the fluctuating Boltzmann equation, as the consequence of the symmetry relations here established. 

These results concern in particular the numerical simulation of dilute or rarefied gases by methods such as the Direct Simulation Monte Carlo method if the enunciated assumptions and, especially, the properties of reciprocity (\ref{reciprocity-sym}) and (\ref{p-reciprocity}) are satisfied \cite{C00}.

At equilibrium where the affinities vanish, the fluctuation relation (\ref{FTC}) reduces to the expression of the principle of detailed balancing, according to which the probabilities of opposite fluctuations are equal.  Out of equilibrium, a bias appears between these probabilities.  Since the ratio of the probabilities of opposite fluctuations goes exponentially with time and the affinites, one of both probabilities may soon become so tiny that it would appear negligible, as nonequilibrium constraints increase.  This is the case in regimes very far from equilibrium where the fluxes tend to become unidirectional.  In extreme regimes for instance with shock waves, the fluctuation relation thus expresses the quasi full irreversibility of the flow due to the extreme rarity of fluctuations opposite to the average fluxes.  It is only relatively close to equilibrium that the opposite fluctuations are not negligible.

In the present paper, the results have been obtained for gases composed of a single atomic species.  However, the results extend to polyatomic molecular species, gas mixtures, as well as reacting dilute or rarefied gases, as long as the property of microreversibility may be assumed \cite{HCB54,B09}.  The scope opened by these results is very broad since we may also envisage their extension to gases ruled by the relativistic Boltzmann equation or to gases of bosons or fermions ruled by the quantum Boltzmann equations \cite{N28,UU33}.

\begin{acknowledgments}
The main result of this paper has been presented at the conference ``Boltzmann equation: mathematics, modeling and simulations" in memory of Carlo Cercignani at the Henri Poincar\'e Institute, Paris, February 9-11, 2011. This research has been financially supported by the Belgian Federal Government (IAP project ``NOSY").
\end{acknowledgments}


\appendix

\section{Time-reserval symmetry of the binary-collision operator}
\label{AppA}

In this appendix, the time-reversal symmetry relation (\ref{sym-rel-C}) is proved for the binary-collision operator
\be
\hat L^{\rm C}\Phi = \sum_{\lambda\mu\rho\sigma} W_{\lambda\mu\rho\sigma}^{\rm C}\left( \hat E_{\lambda}^{-1}\hat E_{\mu}^{-1}\hat E_{\rho}^{+1}\hat E_{\sigma}^{+1}-1\right) N_{\rho} \, N_{\sigma} \, \Phi
\label{L^C}
\ee
Using Eq.~(\ref{E-dag}), its adjoint is given by
\be
\hat L^{\rm C\dagger}\Phi = \sum_{\lambda\mu\rho\sigma} W_{\lambda\mu\rho\sigma}^{\rm C} \, N_{\rho} \, N_{\sigma}\left(\hat E_{\rho}^{-1}\hat E_{\sigma}^{-1} \hat E_{\lambda}^{+1}\hat E_{\mu}^{+1}-1\right) \Phi
\label{L^C_adjoint}
\ee
Using the time-reversal operator (\ref{Theta}) and Eqs.~(\ref{E--eta})-(\ref{E+-eta}),
the expression in the left-hand side of Eq.~(\ref{sym-rel-C}) is transformed as follows:
\bea
\eta^{-1} \Theta \, \hat L^{\rm C} \left( \Theta \, \eta \, \Phi\right) &=& \eta^{-1} 
\sum_{\lambda\mu\rho\sigma} W_{\lambda\mu\rho\sigma}^{\rm C} \, \hat E_{\lambda^{\rm T}}^{-1}\hat E_{\mu^{\rm T}}^{-1}\hat E_{\rho^{\rm T}}^{+1}\hat E_{\sigma^{\rm T}}^{+1} \, N_{\rho^{\rm T}} \, N_{\sigma^{\rm T}}\, \eta\,  \Phi - \sum_{\lambda\mu\rho\sigma} W_{\lambda\mu\rho\sigma}^{\rm C} \, N_{\rho^{\rm T}} \, N_{\sigma^{\rm T}}\,\Phi \nonumber\\
&=& \sum_{\lambda\mu\rho\sigma} W_{\lambda\mu\rho\sigma}^{\rm C} \, \frac{\langle N_{\rho^{\rm T}}\rangle_{\rm R}\langle N_{\sigma^{\rm T}}\rangle_{\rm R}}{\langle N_{\lambda^{\rm T}}\rangle_{\rm R}\langle N_{\mu^{\rm T}}\rangle_{\rm R}}\, N_{\lambda^{\rm T}} \, N_{\mu^{\rm T}}\, \hat E_{\lambda^{\rm T}}^{-1}\hat E_{\mu^{\rm T}}^{-1}\hat E_{\rho^{\rm T}}^{+1}\hat E_{\sigma^{\rm T}}^{+1} \, \Phi - \sum_{\lambda\mu\rho\sigma} W_{\lambda\mu\rho\sigma}^{\rm C} \, N_{\rho^{\rm T}} \, N_{\sigma^{\rm T}}\,\Phi 
\eea
Now, summing over all the cells $\alpha$ gives the same result as summing over the time-reversed cells $\alpha'^{\rm T}$, so that we get
\be
\eta^{-1} \Theta \, \hat L^{\rm C} \left( \Theta \, \eta \, \Phi\right) = \sum_{\lambda\mu\rho\sigma} W_{\rho^{\rm T}\sigma^{\rm T}\lambda^{\rm T}\mu^{\rm T}}^{\rm C} \, \frac{\langle N_{\lambda}\rangle_{\rm R}\langle N_{\mu}\rangle_{\rm R}}{\langle N_{\rho}\rangle_{\rm R}\langle N_{\sigma}\rangle_{\rm R}}\, N_{\rho} \, N_{\sigma}\, \hat E_{\rho}^{-1}\hat E_{\sigma}^{-1}\hat E_{\lambda}^{+1}\hat E_{\mu}^{+1} \, \Phi - \sum_{\lambda\mu\rho\sigma} W_{\rho^{\rm T}\sigma^{\rm T}\lambda^{\rm T}\mu^{\rm T}}^{\rm C} \, N_{\lambda} \, N_{\mu}\,\Phi 
\ee
By using the property of reciprocity (\ref{reciprocity-WC}) with respect to the equilibrium state of the right-hand reservoir, the first term becomes
\be
\eta^{-1} \Theta \, \hat L^{\rm C} \left( \Theta \, \eta \, \Phi\right) = \sum_{\lambda\mu\rho\sigma} W_{\lambda\mu\rho\sigma}^{\rm C}\, N_{\rho} \, N_{\sigma}\, \hat E_{\rho}^{-1}\hat E_{\sigma}^{-1}\hat E_{\lambda}^{+1}\hat E_{\mu}^{+1} \, \Phi - \sum_{\lambda\mu\rho\sigma} W_{\rho^{\rm T}\sigma^{\rm T}\lambda^{\rm T}\mu^{\rm T}}^{\rm C} \, N_{\lambda} \, N_{\mu}\,\Phi 
\ee
By using Eqs.~(\ref{T-sym-WC})-(\ref{PT-sym-WC}) to transform the second term, we finally obtain
\be
\eta^{-1} \Theta \, \hat L^{\rm C} \left( \Theta \, \eta \, \Phi\right) = \sum_{\lambda\mu\rho\sigma} W_{\lambda\mu\rho\sigma}^{\rm C}\, N_{\rho} \, N_{\sigma}\left( \hat E_{\rho}^{-1}\hat E_{\sigma}^{-1}\hat E_{\lambda}^{+1}\hat E_{\mu}^{+1}-1\right) \, \Phi 
\ee
which is identical to the adjoint operator (\ref{L^C_adjoint}). Q. E. D.


\end{document}